\documentclass[12pt]{article} 
\usepackage{epsfig, graphicx}
\usepackage{psfrag} 
\usepackage{amsfonts}

\newcommand{\be}{\begin{equation}}  
\newcommand{\ee}{\end{equation}}  
\newcommand{\bea}{\begin{eqnarray}}     
\newcommand{\eea}{\end{eqnarray}}

\def\lsim{\mathrel{\rlap{\lower4pt\hbox{\hskip1pt$\sim$}}
    \raise1pt\hbox{$<$}}}                
\def\gsim{\mathrel{\rlap{\lower4pt\hbox{\hskip1pt$\sim$}}
    \raise1pt\hbox{$>$}}}                

\begin{document}
\begin{titlepage}
\begin{center}
{\Large \bf Transitions Between de Sitter Minima}

\vspace{40pt}

{\large Puneet Batra$^a$ and Matthew Kleban$^b$}

\vspace{12pt}
{$^a$  
Department of Physics\\ Columbia University\\
538 W. 120th St. \\ New York, NY 10027, USA }

\vspace{12pt}

{$^b$ CCPP\\ 
Department of Physics\\ New York University\\
4 Washington Pl.\\ New York, NY 10003, USA }

\end{center}

\vspace{20pt}

\begin{abstract}
We investigate the cosmological signatures of instantons mediating
tunneling between de Sitter minima.  For generic potentials the
Coleman-de Luccia instanton does not necessarily exist; when it does
not, the instanton which contributes to the decay rate is the trivial
constant solution, known as the Hawking-Moss instanton.  With the aid
of a toy model we interpret this solution and describe the resulting
cosmology.  In neither the Coleman-de Luccia nor Hawking-Moss case can
the resulting cosmology be closed. An observation of significant positive curvature
 would therefore rule out the possibility that our universe
arose from any transition from a neighboring minimum in the
string-theory landscape.

\end{abstract}
\end{titlepage}

\newpage
\section{Introduction}

In this paper we discuss transitions in theories of scalar fields
coupled to gravity. In particular, we focus on a scalar potential that
has multiple positive energy minima. Our objective is to generalize
the work of \cite{Freivogel:2005vv}, which describes the cosmological
signatures of one particular type of transition: the Coleman-de Luccia
(CdL) bubble \cite{Coleman:1980aw}.  It is known that for some
scalar potentials CdL mediated transitions do not exist---in fact,
they exist only when the potential barrier is narrow relative to its
height in Planck units. For most other potentials, it is the
Hawking-Moss (HM) instanton that dominates the
transition \cite{Hawking:1982my}.

In the context of the string theory landscape it is imperative to
study the characteristics of transitions which end in minima with
small positive cosmological constant \cite{Bousso:2000xa,Kachru:2003aw,Susskind:2003kw}.  Transitions mediated by a CdL
bubble have rather special characteristics, and typically do not lead
to universes that resemble our own.  If our universe resulted from
such a transition one can hope that some signature of the special
initial conditions, and the landscape itself, may remain and be
observable today.  Such signatures include negative curvature and a
power spectrum which deviates from scale invariance at large scales.
An interesting conclusion of \cite{Freivogel:2005vv} was that a
measurement of positive curvature larger than $10^{-5}$ would rule out
CdL transitions in the landscape.  We emphasize that average spatial
curvature is a meaningful quantity only when it is larger than $\delta
\rho / \rho$, because different choices of spatial slicing will change
its value by an amount of that order \cite{Gruzinov:2006nk}.

Here we extend the analysis to the case where the CdL instantons
do not exist. This occurs when the scalar potential is broad and
satisfies $V'' < 4 R^{-2}$, where $R$ is the size of the
metastable de Sitter space.  We will see that cosmologies that
follow from HM instantons again have special characteristics, although
they do not have negative curvature on large scales. When the barrier
is such that no CdL instanton exists, we find that the top of the
barrier is broad enough to support slow-roll eternal inflation, and that the cosmology of
universes in the true vacuum following a Hawking-Moss transition is similar to the cosmology of regions
which descended from slow-roll eternal inflation. These regions are
extremely inhomogeneous and anisotropic on large scales, and the
average curvature is not a well-defined quantity.  A period of
ordinary inflation after the transition will of course produce a smooth region
on intermediate scales, with an average curvature at any scale no larger than the density perturbation at that scale.  At larger scales the
perturbation spectrum will become stronger and the space will deviate
more and more from homogeneity and isotropy.  We find that a
measurement of $| \Omega_k|$ significantly larger than $10^{-5}$ in our
universe rules out an HM transition in our past. It is important to
state clearly that no anthropic estimates are used in this bound, and
the only input requirement is the observed smoothness present in the
Cosmic Microwave Background.

There has been some debate over the interpretation of the HM
instanton.  Before discussing cosmology 
we will first exhibit a simple and intuitive way to classify
all instantons with $O(4)$ symmetry that can contribute to the
transition, discuss the number of negative modes of such solutions,
and clarify the role of the HM instanton in the transition process. We
begin by studying quantum mechanics at finite temperature with a
double-well potential (Section 2).  In this case there is a pattern of
instantons very closely analogous to the case of field theory in de Sitter
space, but the instanton analogous to HM can be given a very clear and precise
interpretation.  In Section 3 we will move to the case of field theory
in de Sitter space.  In Section 4 we will consider the consequences
for the cosmology of regions which appear in the true vacuum due to
the HM instanton.

\section{Thermal Tunneling in Quantum Mechanics - a Toy Model}
\label{sec:qm}

In this section we consider tunneling in quantum mechanics at finite temperature.  As we will see, many of the essential features of the CdL and HM instantons appear in this toy model.

\begin{figure}[t]  
\psfragscanon
\psfrag{Vx}[bl]{\Large{$V(x)$}}
\psfrag{Vf}[c]{$V_-$}
\psfrag{Vt}[c]{$V_+$}
\psfrag{V0}[c]{$V_0$}
\psfrag{xf}[c]{$x_-$}
\psfrag{xt}[c]{$x_+$}
\psfrag{x0}[c]{$x=0$}
\psfrag{Ve}[bl]{\Large{$V_{\rm euc}(x)$}}
\psfrag{Vdp}[c]{$V_{euc}''(0)=\omega^2$}
\centerline{\includegraphics[width=5in]{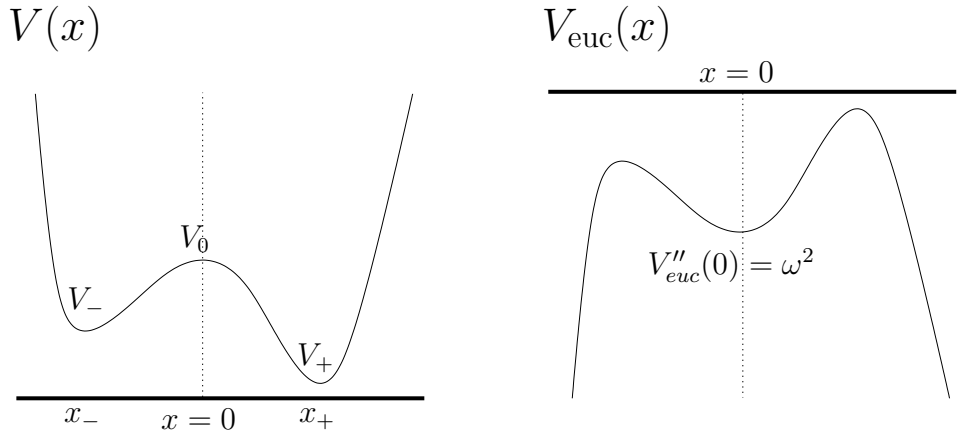}}
\caption{The left-panel shows a potential with an unstable minimum, $V(x-)=V-$. The right-panel shows the upside-down potential $V_{\rm euc} = -V(x)$ used to evaluate the decay rate.}
\label{fig:pot}  
\end{figure} 

To begin, consider the potential in Figure \ref{fig:pot} which has a barrier in between two harmonic wells.  The WKB
approximation can be used to compute the zero-temperature probability of barrier
penetration for a particle with definite energy $E$ in the ``false
vacuum"---the left-hand well.   One computes the action for solutions to the Euclidean equations of motion which begin and
end at the classical turning points of the motion determined by $E$. If these trajectories are unstable (that is, if there is a
small fluctuation around the path with a negative eigenvalue) by
analytic continuation one expects they will contribute an imaginary
part to the energy of the unstable particle \cite{Langer:1967ax,Coleman:1977py}.

However, because Euclidean de Sitter space is compact a better
analogue of the problem at hand is quantum mechanics at finite
temperature $T$.  In that case we study the Euclidean path integral with
time periodically identified, with period $\beta = 1/T$.  Tunneling
probabilities can then be estimated using periodic classical solutions
to the Euclidean equations of motion, and unstable fluctuations around these periodic classical solutions give imaginary corrections to the free energy.\footnote{Recall that the
Euclidean action is $S_E = \int \dot x^2/2 + V(x)$, so solutions to
the Euclidean equations of motion are solutions to an analogue real
time problem with inverted potential $V_{\rm euc} = -V$.}

The finite-temperature partition function is
\be\label{part}
 {\rm Tr} e^{-\beta
H} = \sum_i e^{-E_i \beta} = \int dx~\langle x | e^{-\beta H} | x \rangle = \int dx \int_{x(0)
= x}^{x(\beta)=x} \left[ dx(\tau) \right] e^{-S_E[x(\tau)]}, 
\ee
 where the last expression is the Euclidean path integral with Euclidean time
periodically identified: $\tau \cong \tau + \beta$.

We can begin by preparing the system in a density matrix
describing a collection of particles
localized in the left well, with an energy distribution determined by
the Boltzmann factor $e^{-\beta E_i}$. We consider only temperatures far below the height, $V_0$, of the central barrier.
We are interested in computing an imaginary part of the partition function defined with this new density matrix.  
The reader might be concerned about this, since the exact
thermal partition function in Eq. (\ref{part}) is manifestly real.   From our point of view
this is a consequence of detailed balance: at equilibrium the population of particles in the two wells is such that the rate at
which probability leaks through the barrier from left to right is
equal to the rate from right to left, so that there is no net
probability current.  If we write ${\rm Im} (E_i) =
2 \Gamma_i$, the $\Gamma_i$'s are the decay rates for states on one side and determine the probability current.\footnote{This follows by
writing the decay rate for a unit normalized state on one side in terms of the
probability current through the barrier \cite{Bender:1969si}.}  Therefore if we begin out of equilibrium, for
example in the density matrix representing particles only in the
left-hand well, the probability current is non-zero and the partition function has a small imaginary part.

We therefore expect that 
\bea \label{aff} {\rm Im}(Z_\beta) &\simeq&
-\beta \sum {\rm Im}(E_i) e^{-E_i \beta} \simeq {\rm Im} \left( \int
dx \int_{x(0) = x}^{x(\beta)=x} \left[ dx(\tau) \right]
e^{-S_E[x(\tau)]} \right), \label{eq:imz} \\
&\Rightarrow& \frac{1}{Z} \sum {\rm Im}(E_i) e^{-E_i \beta} \simeq {\rm Im}(F),
 \nonumber 
\eea
where $F$ is the free-energy of the system and we have assumed that the imaginary part of the energies $E_i$ are
small.  Therefore the total decay rate, defined as the Boltzmann weighted average of decay rates for particles of each energy, is proportional to ${\rm Im}(F)$, which itself can be computed using a saddle point approximation, and we see that
 ${\rm Im}(F)$ receives contributions from the decay of all the particles in the thermal distribution rather than a single one.

The $ {\rm Im}(E_i)$'s on the left hand side of Eq. (\ref{aff}) can be defined either by the right hand side
or via the WKB approximation for a single particle with energy $E_i$.  Indeed  Affleck has demonstrated this equivalence
explicitly (for temperatures much less than the barrier height) by computing the ${\rm Im}E_i$'s using the WKB approximation and evaluating both sides \cite{Affleck:1980ac}.  In the case that the temperature is large compared to the second derivative of the potential at the top of the barrier (which is the case where the ``HM" instanton dominates) states with energy slightly above the barrier become important.  In this case one must be careful about using WKB to compute an imaginary part for the energy, and this leads to a logarithmic correction to the formula which will not be important for our purposes.

\subsection{Euclidean solutions}
We will now compute the right-hand side of Eq. 
(\ref{aff}) by saddle point approximation.  If the potential expanded around the maximum has a non-zero
quadratic term, $V_{euc}(x) = -V(0) + (1/2)\omega^2 x^2 + {\cal O} (x^3)$,
then the Euclidean solutions will have a minimum period of oscillation $ 2 \pi
/ \omega$ around the minimum of the Euclidean potential.  When this minimum period is longer than the allowed period $\beta = 1/T$ in the thermal path integral,  $2 \pi / \omega > \beta  $, there are no
oscillating solutions to the equations of motion.  This leaves only the two
stable solutions $x=x_\pm$ and the unstable ``HM" solution
$x=0$.  

\begin{figure}[t] 
\begin{center} 
\psfragscanon
\psfrag{beta}[c]{\Large{$\cal T$}}
\psfrag{bet1}[c]{$2 \pi \over \omega$}
\psfrag{bet2}[c]{$4 \pi \over \omega$}
\psfrag{bet3}[c]{$6 \pi \over \omega$}
\psfrag{x}[c]{\Large{$x$}}
\psfrag{xmax}[c]{$x_{+}$}
\psfrag{zer}[c]{$0$}
\psfrag{1bou}[c]{CdL}
\psfrag{2bou}[c]{2 bounces}
\psfrag{3bou}[c]{3 bounces}
\psfrag{HM}[ct]{HM}
$\begin{array}{c@{\hspace{.5in}}c}\includegraphics[width=2.5in]{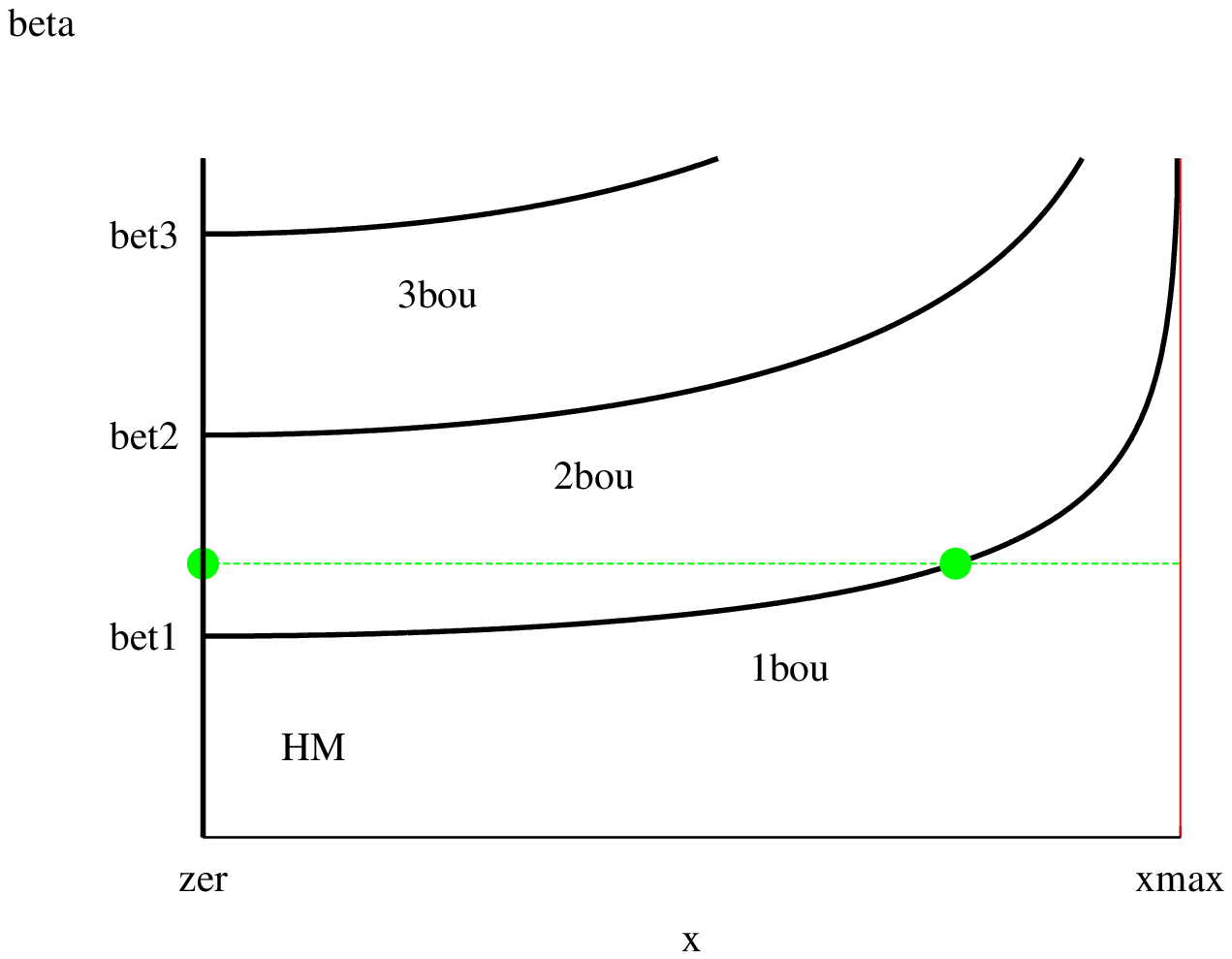} & \includegraphics[width=2.5in]{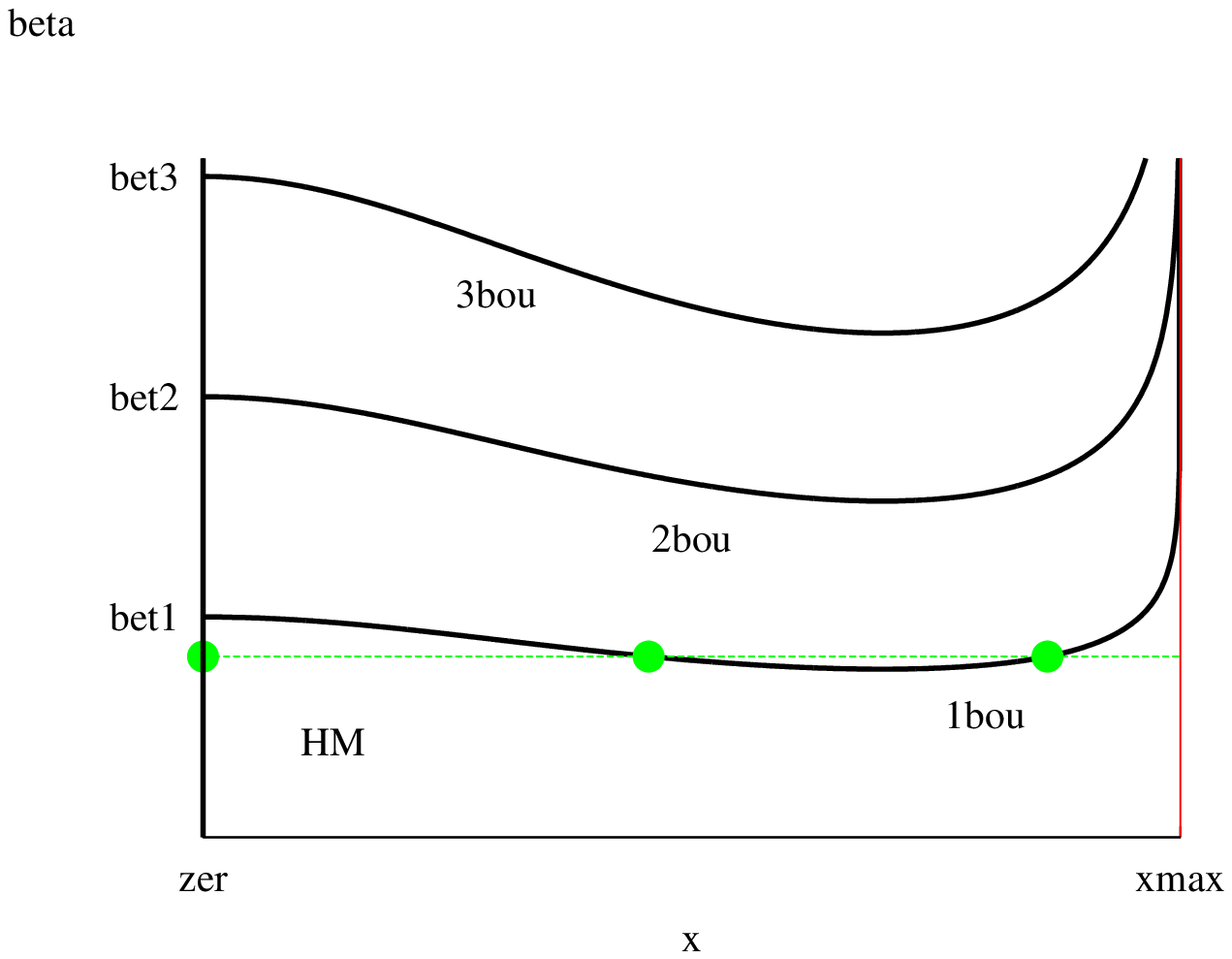} \end{array}$
\end{center}
\caption{Every periodic solution to the euclidean equations of motion is characterized by an amplitude $x$ and a period $\cal T$. Here we plot ${\cal T}(x)$.  The figure on the left is for a potential $V \sim -x^2 + x^4$, and the figure on the right has $V\sim -x^2 -x^4 + x^6$.   To find the set of allowed euclidean solutions for given $\beta$, draw the horizontal line $\cal{T}=\beta$.  The values of $x$ where this line intersects the curves $\cal{T}(\beta)$ are the allowed values of the amplitude.  The line labeled CdL is a classical solution which crosses the maximum of the potential once per $\beta$-period, the ``2 bounces'' solution crosses the maximum twice, etc. The HM solution is the vertical axis $x(\tau)=0$.  A similar figure, based on a communication from one of us, appeared in \cite{Bousso:2006am}.}
\label{fig:bounces}  
\end{figure}

Even when oscillating solutions exist they never touch the false or the true vacuum.  This is
clear from Figure \ref{fig:pot}:  such a solution would take
infinite euclidean time to come to rest on top of the local maximum at $x_-$.  In fact, there is an elegant way to
characterize all Euclidean solutions for potentials with two local
minima. For very small oscillations the
period of oscillation is simply $ 2 \pi / \omega$ (for simplicity we
will assume $\omega^2 = V_{euc}''(0)= -V''(0) \neq 0$, but the extension
to the special case $\omega = 0$ is not difficult).  
Restricting to even powers of $x$ for simplicity, if the sign of the next term in the
power series expansion of the Euclidean potential is positive
(negative), the period for slightly larger oscillations will be
shorter (longer).  However, the amplitude of the oscillation can never
go past the critical points of the potential, since any solution
starting (at rest) there would take infinite time to descend.  Therefore if we plot
the period as a function of amplitude, new zero-amplitude solutions appear when $2 \pi / \omega $
is $ \beta / N$, where $N$ is some integer, and the period must asymptote to infinity as the 
amplitude approaches the critical points of the potential
(see Figure \ref{fig:bounces}).

In the case where the lines on this plot have critical points away from zero amplitude, pairs of solutions appear and disappear at special
values of $\beta$.  Such a pair can appear even in the region below the point $2 \pi / \omega = \beta$, but there must always be a minimum value  $\beta_{\it min}$ below which there are no periodic solutions (since a finite amplitude oscillating solution will always have a finite period).  

\subsection{Negative modes}

In general a saddle point of an integral will contribute an imaginary
part only if there are an odd number of negative modes of the
fluctuation operator around that saddle, because a saddle $x_s(\tau)$
contributes $e^{-S(x_s)}/\sqrt{\det{(S''(x_s))}}$ to the integral.  To
find the number of negative modes of the operator $ \triangle \equiv
S'' = -\partial_{\tau}^2 + V''(x_s)$ we can look for the zero mode, and
then count its nodes.  This determines the number of negative modes by
the usual argument (see e.g. \cite{messiah} p. 109).  This is easy, because given the
solution $x_s(\tau)$ the function $\partial_{\tau}x_s$ is a zero mode of the
fluctuation operator: \be \triangle~ \partial_{\tau}x_s = -\partial_{\tau}^3 x_s + \left(\partial_{\tau}x_s\right) V''(x_s) =
\partial_{\tau} \left( -\partial_{\tau}^2 x_s + V'(x_s) \right) = 0.  \ee A solution
$x_1(\tau)$ which goes through one oscillation per $\beta$-period has
two nodes, as does $\partial_{\tau}x_1$.  Since $\tau$ is compact there are no
solutions with an odd number of nodes, and therefore there is
precisely one negative mode for $x_1$.  By the same argument an
$n$-oscillation solution has $n$ negative modes.

This argument fails for the case of HM solution $x_0$ since $x'_0 = 0$.  However in this case it is trivial to diagonalize $\triangle$: 
\be
{\rm spectrum}(\triangle) = {\rm spectrum}(-\partial_\tau^2 - \omega^2) = \{(2 \pi n/\beta)^2 - \omega^2, n \in {\cal Z}  \}.
 \ee
Therefore the number of negative modes $n$ of the HM instanton is $n = {\rm ceil} \{\beta \omega/(2 \pi) \},$ where ceil means $1 + {\rm integer~part}$.  The appearance of a new negative mode always coincides with the appearance of a new solution.

\subsection{Interpretation}

It is now clear that the imaginary part of the path integral can usually be
approximated by a single saddle point, which is either a solution
$x_1(\tau)$ which bounces once ($2 \pi/ \omega < \beta$) or is the
HM solution $x_0$ ($2 \pi / \omega > \beta $).  Close to the cross-over point there may be multiple instantons contributing (see Figure \ref{fig:bounces}). 

The interpretation of either of these solutions is at first sight
rather unclear.  Neither connects the false vacuum to the true vacuum,
but on the other hand each (in the region where they dominate) has
one unstable mode, and hence should be expected to contribute to the
imaginary part to the vacuum energy expanded around the left-hand
well.

\begin{figure}[t]
\psfragscanon
\psfrag{EN}[ct]{\Large{$ E $}}
\psfrag{EV}[c]{$V_0$}
\psfrag{GE}[c]{\large{$\rho(E) \Gamma(E)$}}
\psfrag{T2}[r]{$2$}
\psfrag{T.995}[l]{$.995 $}
\psfrag{T.2}[l]{$.2$}
\psfrag{.9V0}[l]{{$.9 V_0$}}
\psfrag{1V0}[l]{{$V_0$}}
\psfrag{1.1V0}[l]{{$1.1 V_0$}}
\psfrag{p.05}[l]{$.05$}
\psfrag{p.1}[l]{$.1$}
\psfrag{p.15}[l]{$.15$}
\centerline{\includegraphics[width=3.5in]{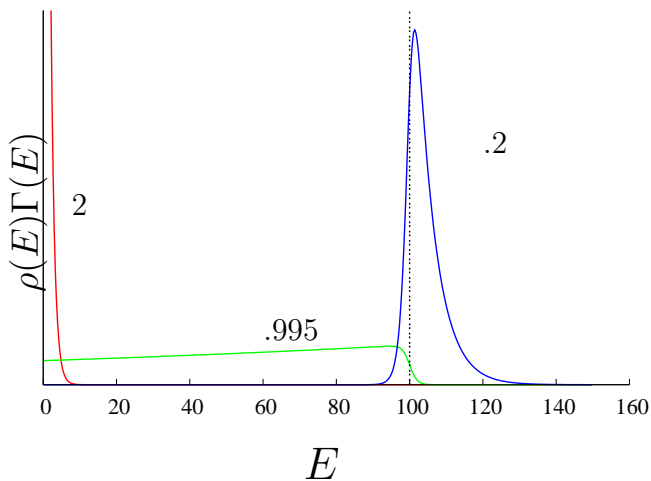}}
\caption{Normalized probability distributions for the thermal decay rate for particles with energy $E$.  Energies are measured in units of $\omega / 2 \pi$.   From left to right, the curves are for $ \beta = \{ 2, .995, .2\} 2 \pi/ \omega$, which correspond to CdL, the transition point, and HM respectively.  The vertical dashed line is $E=V_0=100~ \omega/ 2 \pi$, the energy at the top of the barrier.}
\label{fig:decays}  
\end{figure}

To gain some further intuition let us consider the case where $2 \pi / \omega 
\gg \beta $, but $\beta V_0 \gg 1$.  In this case the
distribution of energies of the particles tunneling through the
barrier is sharply peaked around $E = V_0$.  Physically, even though
the Boltzmann factor suppresses the density of states at higher
energies,  the tunneling rate per particle is much higher (because $\omega$ is small and the barrier quickly broadens as one moves down from the top). This
can be seen explicitly from the form of the distribution in simple cases where
the WKB amplitude can be calculated analytically, for example if we approximate
the barrier as an inverse harmonic oscillator. In this case the
product of the Boltzmann factor and the WKB amplitude are competing
exponentials, 
\be e^{-\beta E}* \Gamma(E) \propto e^{-\frac{2 \pi
V_0}{\omega}} e^{E(-\beta+ \frac{2 \pi}{\omega})} \ee 
where it can be
easily seen that the ratio of $\omega \beta$ determines
whether small energies or large energies contribute most to the
thermal decay rate. Of course for $E > V_0$ the distribution falls
exponentially rapidly because of the Boltzmann factor (no tunneling is necessary for $E > V_0$).  We plot this distribution as a function of 
$\beta$ for $\beta <\pi / \omega$, assuming fixed $\omega$ and $V_0$ and including normalization, in Figure \ref{fig:decays}. 

It is apparent in Figure \ref{fig:decays} that even when the HM
instanton is responsible for the tunneling, some proportion of the particles making it through
the barrier have $E < V_0$. This proportion decreases as the potential is made
broader, since states of lower energy have a harder time
tunneling. Physically, if an observer in the right-hand well measures
the energies of the particles she sees coming occasionally from the
left, she will observe that nearly all of the particle energies are
consistent with an evolution where the particle started at rest and
very close to the top on the right hand side.  The particles in the left-hand well are evaporating.

\section{Thermal Tunneling with Gravity}

In this section we extend our discussion to transitions in the
case of a scalar field theory coupled to gravity, where the original
metastable state is de Sitter space.  We again find that for sufficiently broad potentials only the HM instanton contributes to the decay rate.  This problem has been studied previously in e.g. \cite{Hawking:1982my,Linde:1981zj,Jensen:1983ac,Goncharov:1986ua, Balek:2003uu, Hackworth:2004xb, Hackworth}.

The two quantities of greatest interest are the overall decay rate and the
initial data for the cosmology which results after a transition. In the quantum
mechanics problem, the first is the Boltzmann average of the decay
rates for particles of fixed energy given by the imaginary part of the free energy. The second is either a particle at
rest that appears at the turning point on the other side of the
barrier, as described by the WKB computation, or a particle that appears at the top of the barrier (with some small kinetic energy) after having acquired enough energy from the thermal bath.

In field theory in curved spacetime we lack
the analog of the WKB computation, but we may still use ${\rm Im}(F)$ to describe the rate
of tunneling.  In the CdL phase, the analog of the individual particles tunneling
is the formation of bubbles.  In the HM phase, where the dominant decay channel for particles is that they occasionally acquire enough energy to evaporate over the top, the analog is the random fluctuation of a region of the space to the top of the barrier.
We will discuss this further (and the consequences for cosmology) in the next section.

For concreteness, we imagine a scalar field whose potential $V(\phi)$
is similar to the potential $V(x)$ in Figure \ref{fig:pot}, the
potential with a barrier separating two local minima. We take the
point of view that, in order to compute the decay rate of a
meta-stable de Sitter minimum, we need to consider only the solutions
of the equations of motion for which the topology of the metric is
$S_4$, the Euclidean topology of de Sitter space.  This is an assumption which absent a more complete
theory of quantum gravity we are unable to prove.\footnote{In string
theory (a theory of 2-dimensional quantum gravity over which we have
some control) spaces of differing topology contribute to the partition
function, but the action is proportional to the Euler character times
the coupling constant, so that non-spherical topologies are
subdominant.  It is possible that in higher dimensions as well other
topologies can contribute, but we have no way of determining their
relative action as we do not even have perturbative control over the
path integral.}

It is not difficult to extend the techniques developed in the last
section to the case of field theory on $S_4$.  The decay rates are
again proportional to ${{\rm Im}(F})$, but the calculation is more involved because  
the space of Euclidean solutions is much larger and more difficult to
classify than for the quantum mechanics of a single particle moving in
the same potential.

In what follows we will largely ignore the backreaction of the field
on the metric.  This assumption is good in the limit that the
variations in the scalar potential are much smaller than its
magnitude, $V' / V \ll 1$.  
Note that this condition is identical to one of the slow roll conditions for inflation. 
In this limit, $V_- \sim V_0$, so the radius of the $S_4$
doesn't vary significantly as the field oscillates across the maximum.
In fact, in the regime we are primarily interested in (where $V'' \ll V$ and the HM instanton
is the only solution) we incorporate the back-reaction and treat the problem exactly, since the only effect of back-reaction is that
the sphere radius is determined by $V_0$ rather than $V_-$.

In
general, we do not expect that this assumption will be valid in the
string theory landscape, but on the other hand we also do not expect
backreaction to change our conclusions qualitatively so long as the
metric remains compact.

The Euclidean action is 
\be S = \int_{S_4} \sqrt{g} \left( \frac{1}{2}g^{\mu \nu}
\partial_\mu \phi \partial_\nu \phi + V(\phi) \right).  \ee 
To gain some insight, we can begin by solving this equation in the case where $V_{\rm euc}$ is harmonic.  As in the quantum mechanical toy model the solutions oscillate about $\phi=0$.  The eigenvalues of the Laplace operator on an
$S_d$ of radius $R$ are 
\be
\lambda_l=l(l+d-1)/R^2
\ee
with multiplicity 
\be\label{mult}
N_l = {(2l + d - 1)(l+d-2)! \over l! (d-1)!},
\ee
for $l \in \{0,1,2,3, \ldots\}$. Oscillating solutions exist only for certain values of $\omega$:  they exist iff $-V''(\phi=0) \equiv \omega^2= l(l+d-1)/R^2$ with $l > 0$. 
As we will see below, in a non-harmonic potential this condition determines the existence of small amplitude fluctuations in a way totally analogous to the finite temperature quantum mechanics (see Figure \ref{fig:bounces}).
Here the inverse radius of the $S_4$ is $R = H^{-1} \sim 
V_-^{-1/2}$ in units where  $8 \pi G / 3 \equiv M_{\rm pl}^{-2}$.  The existence of periodic euclidean solutions
depends on the potential, and only the constant $l=0$ solution exists if $V'' < 4/R^2 = 4H^2$.  This solution is known as the
Hawking-Moss instanton \cite{Hawking:1982my}.

To proceed beyond the small amplitude approximation, we will assume that the most symmetric
solutions--those with at least an $SO(4)$ invariance, so that the
field configurations depend only on a single angle--have the smallest
action and hence dominate the path integral.  This assumption is well
motivated: it was proven for field theory in $d=4$ 
\cite{Coleman:1977th} and later when the coupling to gravity is
included \cite{Tanaka:1999pj}.  Explicit
constructions of less symmetric solutions show that they have larger action and therefore support this
conclusion \cite{Garriga:2004nm}.\footnote{As we will see there will in general be many
solutions invariant under $SO(4)$ (for example solutions with multiple
bounces), and it is only the member of this set with the least
action---which is either a single bounce CdL or a HM solution---that we will need to assume is the dominant one.  The rest, in general,
may have larger action than some of the less symmetric solutions.}

With this assumption, the field configurations and metric depend only
on a single coordinate and the problem reduces to a quantum mechanics,
in which case the analysis of the previous section goes through
essentially unchanged.  The only significant difference is the
presence of additional zero modes which are dealt with in the standard
way \cite{Coleman:1977py} and the presence of a
friction term in the equations of motion.

We will therefore look for $SO(4)$ invariant solutions to the equations of
motion in a potential of the form of Figure \ref{fig:pot}. As usual,
the Euclidean field equations can be thought of as the motion of a
particle in a potential $V_{\rm euc}$.  We ignore backreaction and
take the metric to be $ds^2 = d \theta^2 + \sin^2\theta~d \Omega_3^2$.
Field configurations $\phi = \phi(\theta)$ are then invariant under
the $SO(4) \subset SO(5)$ which rotates the $S_3$. Under these assumptions we can write down the equations of motion
 \be\label{eom}
{1 \over \sqrt{g}} \partial_\mu \left( \sqrt{g} g^{\mu \nu}
\partial_\nu \phi \right) = \frac{1}{R^2} \left[\phi''(\theta) + 3
\cot\theta~\phi'(\theta)\right] =V'(\phi).
\ee 
Each solution begins at some
value $\phi(\theta =0)$ at the north pole of the $S_4$ with
$\phi'(0)=0$, and has $\phi'(\pi)=0$ at the south pole. Since the
equation is reversible under $\theta \rightarrow \pi - \theta$, there
is always a 'partner solution' that travels in the opposite direction,
but crosses the barrier the same number of times for any solution that
has non-zero amplitude.  For potentials in the class illustrated in Figure
\ref{fig:pot}, three constant solutions are immediately obvious: $\phi
= 0, \phi=\phi_+$, and $\phi = \phi_-$.  Of these, only $\phi=0$ (the HM instanton) could
contribute to a decay amplitude, as it is the only one which might
have a negative fluctuation mode. 

After our analysis of quantum mechanics at finite temperature it is clear that the
HM solution always has at least one unstable mode.  In the
limit that the barrier is very broad, there is only one, since as
before the spectrum of the fluctuation operator $\triangle = -\Box +
V''$ always has a lowest eigenvalue $\lambda_0 = V''(\phi=0) < 0$.
Indeed, in general the spectrum of scalar modes expanded around a
constant solution $\phi = \phi_0$ on the sphere is simply 
\be
\lambda_l = {l(l+3) \over R^2} + V'' = {l(l+3)} V(\phi_0) +
V''(\phi_0), 
\ee
 and $N_l$ (see Eq. (\ref{mult})) new negative modes of the instanton appear when
 $l(l+3) V(\phi_0) +
V''(\phi_0) = 0$ for some $l$.  This conclusion for the number of negative modes of the
solution is unaffected by backreaction, since the
fluctuations are by definition taken to quadratic order and therefore
the background is precisely $S_4$. One might wonder whether fluctuations in the additional degrees
of freedom coming from the metric change this counting.  However if the
field theory limit ($M_{\rm Pl} \rightarrow \infty$) is to be smooth this must not
be the case, and therefore when $H/M_{\rm Pl} \ll 1$ (and effective field theory
applies) there should be no additional negative modes.  In addition, it has been argued on general grounds that no additional
negative eigenvalues appear when gravity is included \cite{Tanaka:1999pj,Gratton:2000fj}.

The next step is to consider when new small amplitude solutions appear.
Backreaction is irrelevant for this consideration for the same reason
as above--since the new solutions have zero amplitude when they
appear, they can not affect the geometry.  Therefore the condition is
again as it was for quantum mechanics: new solutions emerge from
HM precisely at the point new negative modes appear, namely when
 $V''(\phi=0) = l(l+3)/R^2$.

 The last difference is the presence of the friction term in Eq. (\ref{eom}), which means
that a solution with $n$ bounces with period $\beta$ is not also periodic with period $\beta/n$, as is the case in quantum mechanics.  This is due to the friction term--the amplitude of successive bounces
decreases monotonically as $\theta$ increases from zero, and then
increases monotonically when $\theta>\pi/2$.  Therefore the maximum
amplitude is always the value at $\theta = 0$ or $\pi$, and so for the
purposes of Figure \ref{fig:bounces} the amplitude plotted is the
value of the field at the pole. 

\section{Cosmological signatures}

In this section we consider the consequences of these transitions
for cosmology in the true vacuum.  There are three cases of interest:
when the potential barrier is very narrow compared to its height, very
broad, and the region in between.  In the first case CdL
instantons dominate, in the second HM, and then last depends on the
details of the potential and is difficult to analyze.

\subsection{Narrow Barrier}

In the case the barrier is narrow a single-bounce instanton with
$O(4)$ symmetry exists.  This is the case analyzed in
\cite{Freivogel:2005vv}.  Part of the region inside the bubble forms
an open FRW cosmology, but with a curvature dominated and non-singular
``big-bang."  The initial conditions at the big bang are that the
field is at rest and takes the value $\phi_+$--the value of the field
at the pole of the sphere on the true vacuum side ($\theta=\pi$ for
example).  Fluctuations around this instanton are small if the overall
scale of the potential is less than $1$, and so the initial condition
shortly after the bang is an extremely homogeneous negatively curved
cosmology.  If the potential is such that the field quickly rolls down
to its true minimum without inflating, there is no mechanism to
generate density perturbations or to reduce the large curvature
significantly, and the resulting empty, negatively curved universe
expands homogeneously.

In order to produce a universe containing structure, one needs a
mechanism to both reduce the curvature and generate curvature
perturbations.  The most economical method for doing this is a section
of relatively flat, inflationary potential below the point where the
field appears after tunneling.  If this plateau is sufficiently long
to provide at least 60 efolds of inflation a universe consistent with
current observations will result. A measurement of positive curvature larger than $10^{-5}$
would rule out CdL transitions in the landscape, since the initial condition of slow-roll inflation is a negatively curved space. Interestingly, the minimum number of
efolds necessary to simply produce structure---irrespective of
consistency with observations---is only a few efolds less than the number required to inflate the curvature down to the point where it is consistent with current observational limits \cite{Freivogel:2005vv}. 

Regions of almost flat potential are presumably quite rare in the
string theory landscape, and this is a fine-tuning above and beyond
the tuning of the vacuum energy at the minimum.  Furthermore the
conditions for slow roll $V' \ll V$ and $V'' \ll V$ are in tension with the requirement that the barrier be narrow.  Therefore it might be the
case that long inflation after CdL tunneling is strongly disfavored.
In that case one expects a relatively short period of inflation after
the tunneling and it is reasonable to hope that some observable
signature, such as negative curvature, may remain.

\subsection{Broad Barrier}

The HM
instanton dominates when the barrier is broad and flat, which are conditions suitable for slow-roll eternal inflation on top of the
barrier.  The standard condition for eternal inflation is that the
``quantum velocity"---the root-mean-square deviation in $\phi$ after
one Hubble time---should be larger than the change in $\phi$ due to the
classical velocity after the same time \cite{Linde:1986fd}.  Quantitatively we require
that 
\be\label{et}
 \dot{\phi}/ H < \delta \phi ~~\Rightarrow ~~
V^{3/2}/V' > 2 \pi/3, 
\ee 
using the slow-roll equation of motion $3 H
\dot{\phi} = -V'(\phi)$ and $\delta \phi = H/2 \pi$.  \footnote{Note that
while this condition is equivalent to requiring $\delta \rho/\rho >1$
after the end of inflation, it does {\em not} imply that the local quantum mechanical
fluctuations are large {\em during} inflation--indeed they are simply
of order $\delta \rho \sim V' H \ll H^2 \sim \rho$.}  Obviously this
condition is satisfied at $\phi=0$ where $V'=0$.  However, this is
not sufficient to guarantee eternal inflation if the barrier
is sharp.  Instead, we require that Eq. (\ref{et}) be
satisfied over a region in field space of order $\delta \phi$.  If the barrier
is quadratic close to the maximum this amounts to requiring that $3
V_0/\omega^2 >1$, a slightly weaker condition than for HM to dominate.  
The implication is that potentials for which HM dominates can support
eternally inflating domain walls separating the two vacua, and one can view
the HM instanton, rather than as a transition between the minima, as a process where part of the false vacuum region attaches itself to the wall.  See \cite{Vilenkin:1994pv, Linde:1994wt} for  discussions of 
eternally inflating domain walls.

Therefore in much of the regime where the HM instanton dominates,
regions of space which tunnel to the top of the barrier will undergo
eternal inflation. A region of space which tunneled via the HM instanton
will be identical to any other which descended from an eternally
inflating region of the potential, at least as far as a semi-classical
analysis can show. 

One may wonder whether the HM instanton indicates that the entire
universe tunnels to the top of the barrier.  Indeed, this is the
interpretation taken in the original paper \cite{Hawking:1982my}.
However, our analysis of finite-temperature quantum mechanics shows
that this is not a reasonable assumption.  The analogous statement in quantum mechanics is that all of the
particles in the thermal distribution in the left-hand well jump
simultaneously to the top, an event that not only violates
conservation of energy (since the temperature is much below the
barrier height) but common sense as well.  Instead, the quantum
mechanical ``HM" instanton computes the flux of probability current
correctly, and furthermore indicates that most of the particles which
appear in the right-hand well have an energy close to the barrier
height.  Therefore we conclude that the HM action determines the rate
per unit time per unit volume for a {\em Hubble-sized region} to
tunnel to the top.

There are a number of other arguments supporting this conclusion.
Consider a Hubble region of a weakly curved inflating universe.  The
physics of that region should be independent of the global structure
of the spacetime.  If for example the region were embedded in a flat
or open inflating universe rather than a closed one, the action for
the entire universe to tunnel would be infinite (due to the infinite
volume) and so this process would not contribute.  Therefore this can
not be the correct interpretation of the HM instanton.

More evidence comes from the stochastic approach to tunneling
~\cite{Starobinsky:1986fx, Linde:1991sk}, which is valid in the regime
where $V'' \ll V$.  In this approach the quantum fluctuations are
replaced with a random noise term which causes the field to fluctuate
and occasionally drives small regions to the top of the barrier.  The
decay rates computed this way agree with the HM action, and describe regions of at least size $H^{-1}$ at the top of the barrier.

A final point is that, since the top of the potential is suitable for
eternal inflation in the region where HM dominates, the details of the
regions which tunnel there are not particularly important for the
cosmology seen by an observer in the true minimum.

\subsubsection*{Hawking-Moss cosmology}

We now address the cosmology of an observer who finds herself in the
right-hand well after a transition mediated by the HM
instanton. Let us first assume that no period of ordinary inflation occurs
after exiting the eternally inflating regime (defined as the regime in which $V^{3/2}/V' > 1$). If we inhabit a patch of the
eternally inflating space which has descended and reheated in the the true minimum,
Eq. (\ref{et}) shows that $\delta \rho/\rho$ will be of $\cal{O}$$(1)$
 \cite{Linde:2005ht}.  The
universe would be very inhomogeneous and look nothing like the
one we live in, and in fact might not even be suitable for structure
formation \cite{Tegmark:1997in}.

To recreate the observed spectrum of density fluctuations we must
instead assume that the potential is such that there was a significant period of ordinary
slow-roll inflation on the true-vacuum side of the potential
barrier below the eternally inflating regime. The $O(1)$ density perturbations from the
period of eternal inflation will be pushed out to exponentially large
scales by this later period of expansion, and replaced with smaller $\delta \rho/
\rho \sim V^{3/2}/ V'$, which may become small as $V'$ increases.  For example, suppose $N$ efolds are required to blow up a single Hubble volume at an initial time $t_N$ into our current observable universe ($N \sim 60$ depending on details of the reheating
temperature, post-inflation expansion history, etc.). This would predict that the curvature $\Omega_{k_N}$ of a Hubble patch a time $t_N$ corresponding to
$N$ efolds before the end of ordinary inflation should be equal to $\Omega_{k_0}$, the
curvature measured today.

If this primordial patch had large density perturbations (for instance if it was in the eternally inflating regime) our universe would not be close to an FRW universe on large scales today---the CMB would not be homogeneous or isotropic, and in this case spatial curvature on large scales is not a meaningful concept because there is no preferred FRW slicing.  If instead some number of efolds $M$ of ordinary inflation took place after the exit from eternal inflation and before $t_N$ the spatial curvature on some slice would have been reduced by a factor of $e^{2M}$.  At the same time the initial large density perturbations on that slice would also be inflated away, and so the patch would look approximately flat and homogeneous (up to any additional density perturbations being generated through quantum fluctuations).  We always bear in mind that any quantity which depends on a choice of FRW slicing is well-defined only up to an uncertainty of order $\delta \rho / \rho$.

Therefore if some period of ordinary inflation took place after the exit from the eternally inflating regime, the universe will look close to flat on small scales,
with fluctuations in the curvature of order the fluctuations in the
density for a generic choice of spatial slice.  On larger scales, corresponding to values of $\phi$
approaching the top of the barrier, the density perturbations will
grow and the average curvature will become less and less well defined.
Finally at very large scales the universe will simply not be homogeneous
and spatial curvature ceases to be a useful concept.
In the simplest cases the prediction is then that the spectrum of
$\delta \rho/\rho$ is red, and the measured curvature is less than or
of order $\delta \rho/\rho$.  A measurement of average curvature
significantly different from $\delta \rho/\rho$ would rule out the
model.  In particular if our universe originated in a Hawking-Moss tunneling,  
\be
|\Omega_{k_0}| \lsim 10^{-5}.
\ee

Another prediction that distinguishes this type of tunneling from CdL
are tensor modes.  In CdL tunneling the tensor-to-scalar
ratio goes to infinity on scales where the curvature becomes important
(in other words scales corresponding to values of the field close to
the point where it emerged after tunneling---see for
example~\cite{Hawking:2000ee}).  In our case, the tensor to scalar
ratio will go to zero at large scales.  the reason for this is that
the tensor power is proportional to $H$, which increases only
gradually as the field moves up towards the maximum, whereas the
scalar power gets large: \be P_T/P_S \sim H / (V^{3/2}/V') \sim V'/V.
\ee

Therefore there is a clear way to distinguish between the two
possibilities: a measurement of negative curvature, accompanied by an
increase in tensor power at large scales would indicate a CdL
tunneling, whereas an over all increase in $\delta \rho/\rho$, no
curvature, and no or low tensors at large scales indicates HM
tunneling.  A measurement of positive curvature would disconfirm both
models.

\subsection{In between}

In the case that the barrier is neither very sharp nor very broad, the
analysis becomes considerably more complicated.  In this region there
may be an HM instanton and one or more single-bounce CdL type
instantons, all with one single negative mode (c.f. Figure
\ref{fig:bounces}).  In the quantum mechanics problem, at the
transition where a new small-amplitude bounce appears the distribution
of energies is relatively flat (Figure \ref{fig:decays}), meaning that an
observer on the right will see an equal flux of particles with all
energies below the barrier height.  This situation presumably
describes in the field theory a regime in which some CdL type bubbles
form, and other regions fluctuate to the top of the potential and then
roll down the other side.  Determining the precise characteristics of the
resulting cosmology will require further study, but it seems highly unlikely that such a process can produce significant positive curvature.

\section{Conclusion}
In the end, it appears that there are two distinct regimes for
transitions between de Sitter minima.  In the limit that the barrier
is sharp the tunneling proceeds via thin-wall bubble nucleation, which
produces empty, negatively curved FRW cosmologies with very particular
initial conditions, which guarantee among other things that the ``big
bang" is curvature dominated and regular.  These initial conditions
are quite inhospitable to structure formation.  To produce a universe
like ours then requires a significant number of efoldings of inflation
after the tunneling, which both inflates away the curvature and
produces density perturbations.  If large numbers of efoldings are
sharply enough disfavored in the landscape, we may hope to see some
signatures of these initial conditions in the large scale structure of
our universe.

On the other hand if the barrier is broad, the HM instanton dominates.
If our universe originated from such a transition it will be more and
more inhomogeneous on larger and larger scales.  Such a universe does
not posses a well-defined average spatial curvature, because the
average will diverge as the volume it is taken over increases and will
depend sensitively on an arbitrary choice of spatial slicing.
Cosmology in such a universe will be very unlike ours unless a
significant number of efoldings of inflation occur after the descent
from the eternally inflating regime, in which case there will be a
range of scales over which the universe looks approximately flat and
homogeneous.  Again, we may hope to see signatures of the high-scale
initial conditions if the number of efoldings of inflation was not too
large.

One may wonder which of the two possibilities is the more generic in the landscape.  At present we are not able to answer this question, but some analysis is instructive.  The condition for HM to dominate is that 
\be\label{eq:HM}
V/M_{\rm Pl}^2 > V'',
\ee
where we have restored the dimensions and dropped numerical factors.  Since for our effective description to be valid we must require $V << M_{\rm Pl}^4$, it appears that satisfying this equation requires significant fine-tuning.  However, we know that a CdL tunneling must be followed by a significant period of inflation in order to form structure, so that the potential must be quite flat after the barrier.  In fact the condition for slow-roll inflation is identical to the condition in Eq. (\ref{eq:HM}).  Therefore it is not at all clear whether CdL tunneling followed by long inflation is more or less fine-tuned than HM tunneling followed by long inflation.  Understanding this issue will require understanding the measure on the number of efolds $N$.

One thing we can say for sure is that neither of these transitions will produce closed FRW cosmologies.  While such solutions might be possible in the landscape---by fiat one could choose initial conditions that start the universe off with a closed bang in some metastable minimum---they are not comprehensible from the point of view that the landscape was populated from a generic high-scale initial state.  In other words one expects that in the early universe, when the temperature was very high, most or all minima were populated as the universe cooled.  Vacua with large amounts of vacuum energy would then inflate the fastest and dominate the measure, and tunneling events of the type described here would populate the rest of the minima.  On the other hand beginning the universe with a closed bang in one of the minima with very small cosmological constant requires an enormous fine-tuning and seems completely unnatural. 
Therefore in our view a measurement of positive curvature would constitute a dis-confirmation of the landscape.

\subsection*{Acknowledgments}
No anthropics were harmed during the writing of this paper. The authors would like to thank B. Freivogel, G. Gabadadze, A. Gruzinov, T. Hertog, N. Kaloper, J. Kratochvil,  A. Linde, M. Porrati, R. Rabadan, S. Shenker, P. Steinhardt, A. Vilenkin, E. Weinberg, and S. Winitzki for useful discussions,  and especially L. Susskind for insightful conversations throughout course of this work. P.B. thanks the Aspen Center for Physics for their hospitality during a productive stage of this work. PB is supported by the DOE under contract DE-FG02-92ER.

\end{document}